
%
\input harvmac
%
%
%
%
\ifx\answ\bigans
\else
\output={
  \almostshipout{\leftline{\vbox{\pagebody\makefootline}}}\advancepageno
}
\fi
%
%
%

%
%

%
%
\def\UCSD#1#2{\noindent#1\hfill #2%
\supereject\global\hsize=\hsbody%
\footline={\hss\tenrm\folio\hss}}
%
%
\def\abstract#1{\centerline{\bf Abstract}\nobreak\medskip\nobreak\par #1}
%
%
%
%
\edef\tfontsize{ scaled\magstep3}
 \tfontsize  \tfontsize
 \tfontsize \font\titlei=cmmi10 \tfontsize
\font\titleis=cmmi7 \tfontsize \font\titleiss=cmmi5 \tfontsize
\font\titlesy=cmsy10 \tfontsize \font\titlesys=cmsy7 \tfontsize
\font\titlesyss=cmsy5 \tfontsize  \tfontsize
\skewchar\titlei='177 \skewchar\titleis='177 \skewchar\titleiss='177
\skewchar\titlesy='60 \skewchar\titlesys='60 \skewchar\titlesyss='60
%
%
%
%
%
\def\inv{^{\raise.15ex\hbox{${\scriptscriptstyle -}$}\kern-.05em 1}}
\def\lbar{{\lower.35ex\hbox{$\mathchar'26$}\mkern-10mu\lambda}} 

%
%
%
%
\def\slash#1{\rlap{$#1$}/} 
\def\dsl{\,\raise.15ex\hbox{/}\mkern-13.5mu D} 
\def\delsl{\raise.15ex\hbox{/}\kern-.57em\partial}
\def\Ksl{\hbox{/\kern-.6000em\rm K}}
\def\Asl{\hbox{/\kern-.6500em \rm A}}
\def\Dsl{\hbox{/\kern-.6000em\rm D}} 
\def\Qsl{\hbox{/\kern-.6000em\rm Q}}
\def\gradsl{\hbox{/\kern-.6500em$\nabla$}}
%
%
\def\lspace{\ifx\answ\bigans{}\else\qquad\fi}
\def\lbspace{\ifx\answ\bigans{}\else\hskip-.2in\fi} 
%
%
\def\boxeqn#1{\vcenter{\vbox{\hrule\hbox{\vrule\kern3pt\vbox{\kern3pt
        \hbox{${\displaystyle #1}$}\kern3pt}\kern3pt\vrule}\hrule}}}
%
%
\def\mbox#1#2{\vcenter{\hrule \hbox{\vrule height#2in
\kern#1in \vrule} \hrule}}
%
%
%
%
   \def\CD{{\cal D}}
   
   \def\CL{{\cal L}}

%
%
%
%
%

%

\def\bar#1{\overline{#1}}
\def\vev#1{\left\langle #1 \right\rangle}
\def\bra#1{\left\langle #1\right|}
\def\ket#1{\left| #1\right\rangle}
\def\abs#1{\left| #1\right|}

\def\darr#1{\raise1.5ex\hbox{$\leftrightarrow$}\mkern-16.5mu #1}

%
%
\def\half{{\textstyle{1\over2}}} 
\def\frac#1#2{{\textstyle{#1\over #2}}} 
%
%
%
%

\def\Tr{\mathop{\rm Tr}}

\def\MeV{{\rm MeV}}

%
%
%
%

%
%
\def\ltap{\ \raise.3ex\hbox{$<$\kern-.75em\lower1ex\hbox{$\sim$}}\ }
\def\gtap{\ \raise.3ex\hbox{$>$\kern-.75em\lower1ex\hbox{$\sim$}}\ }
\def\gl{\ \raise.5ex\hbox{$>$}\kern-.8em\lower.5ex\hbox{$<$}\ }
\def\roughly#1{\raise.3ex\hbox{$#1$\kern-.75em\lower1ex\hbox{$\sim$}}}
%
%
        \def\etc{\hbox{\it etc.}}
\def\eg{\hbox{\it e.g.}}        
\def\etal{\hbox{\it et al.}}

\def\np#1#2#3{Nucl. Phys. B{#1} (#2) #3}
\def\pl#1#2#3{Phys. Lett. {#1}B (#2) #3}
\def\prl#1#2#3{Phys. Rev. Lett. {#1} (#2) #3}
\def\physrev#1#2#3{Phys. Rev. {#1} (#2) #3}

\relax

\def\bfu#1{{\bf #1}}
\bigskip
\centerline{{\titlefont QCD Mass Inequalities in the Heavy Quark Limit}}
\bigskip
\centerline{Zachary Guralnik}
\centerline{{\sl Department of Physics, University of California at San
Diego}}
\centerline{{\sl 9500 Gilman Drive, La Jolla, CA 92093-0319}}
\medskip
\centerline{Aneesh V. Manohar}
\centerline{{\sl CERN TH-Division, CH-1211 Geneva 23,
Switzerland}\footnote{*}{On leave from the University of California at
San Diego.}}
\vfill
\abstract{QCD inequalities are derived for the masses of mesons and
baryons containing a single heavy quark using the heavy quark effective
field theory.  A rigorous lower bound is obtained for the $\bar\Lambda$
parameters of the heavy quark effective theory that parameterize $1/m$
corrections, $\bar\Lambda\ge237$~MeV for mesons, and $\bar \Lambda \ge
657$~MeV for baryons. The inequalities on $\bar\Lambda$ imply the
inequalities $m_c< 1627$~MeV and $m_b< 5068$~MeV for the mass parameters
of the heavy quark effective field theory.
}
\vfill
\UCSD{\vbox{\hbox{CERN-TH.6766/92}\hbox{UCSD/PTH 92-45}
\hbox{hep-ph/9212289}}}{December 1992}

Rigorous inequalities between hadron masses in QCD can be derived using
the Euclidean functional integral formulation of the theory\foot{A good
discussion of QCD inequalities can be found in the Caltech lecture notes
for Ph230 of J. Preskill (unpublished).} \ref\weingarten{D. Weingarten,
\prl {51} {1983} {1830}}\ref\vafa{C. Vafa and E. Witten, \np {B234}
{1984} {173}}\ref\witten{E. Witten, \prl {51} {1983}
{2351}}\ref\nussinov{S. Nussinov, \pl {139} {1984} {203}}\ref\espriu{D.
Espriu, M. Gross and J.F. Wheater, \pl{146} {1984} {67}}. The basic idea
is that the functional integral measure for a vector-like theory such as
QCD is real and positive, so that the Cauchy-Schwarz inequality can be
used to derive inequalities on Euclidean correlation functions. The
inequalities among correlation functions imply inequalities among hadron
masses. The usual QCD inequalities hold for arbitrary quark masses, and
so apply also to the heavy quark case \eg, $m(\rho) \ge m(\pi)$
\weingarten\ has the heavy quark analog $m(B^*)\ge m(B)$, \etc~ These
will not be discussed further here. QCD inequalities are derived in this
paper for the heavy quark effective field theory \ref\vsiw{M.B. Voloshin
and M.A. Shifman, Sov. J. Nucl. Phys. 45 (1987) 292, Sov. J. Nucl. Phys.
47 (1988) 511\semi N. Isgur and M.B. Wise, \pl {232} {1989} {113}, \pl
{237} {1990} {527}}\ref\georgi{H. Georgi, \pl {240} {1990}
{447}}\ref\eichten{E. Eichten and B. Hill,  \pl {234} {1990} {511}},
which is a systematic expansion about the infinite quark mass limit. The
quark propagator in the infinite mass limit is the path ordered integral
of the exponential of the vector potential, which is a unitary matrix.
This implies certain inequalities which would not necessarily hold for
light quarks.

In this paper, mass inequalities will be derived for hadrons containing
only a single heavy quark  $Q$. The heavy quark effective theory has the
leading order fermion Lagrangian
\eqn\hqlag{
\CL_v =i\, \bar Q_v \left(v\cdot D\right)Q_v,
}
where $Q_v$ is a Dirac spinor field that annihilates a heavy quark with
velocity $v$, and $D$ is the gauge covariant derivative. The total
Lagrangian is the sum of the heavy quark Lagrangian and the usual QCD
Lagrangian for the light quarks and gluons. (In this paper, light quarks
will refer to quarks with mass $m_q$ that is finite, but not necessarily
small compared to $\Lambda_{\rm QCD}$.) The field $Q_v$ satisfies the
constraint
\eqn\hconst{
Q_v = {1+\slash v\over 2} Q_v.
}
The masses of hadrons in the heavy quark effective theory containing a
single heavy quark $Q$ are $m-m_Q$, where $m$ is the mass of the hadron,
and $m_Q$ is the mass of the heavy quark. The heavy quark mass $m_Q$,
which is defined so that the effective Lagrangian eq.~\hqlag\ has no
residual mass term, is a well defined quantity which is a parameter of
the heavy quark effective theory. A detailed discussion of this issue
can be found in Ref.~\ref\fln{A.F. Falk, M. Neubert, and M.E. Luke,
SLAC-PUB-5771 (1992)}.

The mass inequalities can be easily derived using the continuum
formulation of the theory, provided one uses certain ``obvious''
properties of the products of Dirac delta functions. The same results
can also be obtained using a lattice regulated version of the theory
which avoids ``problems'' with Dirac delta functions, and is the method
used here. The Lorentz frame for the heavy quark theory can be chosen so
that the velocity vector $v$ is $(1,0,0,0)$, before analytically
continuing the theory to Euclidean space. The hypercubic Euclidean
lattice is chosen with edges that are parallel or perpendicular to $v$,
for simplicity. The Euclidean coordinate parallel to $v$ will be denoted
by $t$, and the transverse coordinates will be denoted by ${\bfu x}$.
The lattice spacing is $a$, with $t\equiv n_0 a$ and ${\bfu x}\equiv
{\bfu n} a$. The Euclidean propagator for the heavy quark theory is then
\ref\eichfein{E. Eichten and F. Feinberg, \physrev {D23} {1981} {2724}}
\eqn\eprop{\eqalign{
W(\bfu x,t; \bfu y, s) &= a^{-3}\ {1+\gamma^0\over 2}\ U(\bfu x, t;
\bfu x, t-a) U(\bfu x, t-a; \bfu x, t-2a)
\ldots\cr&\qquad\ldots U(\bfu x, s+2a; \bfu x, s+a) U(\bfu x, s+a; \bfu x, s)
\ {\rm if}\ \bfu x = \bfu y,\ t >s\cr &=0\qquad {\rm otherwise}.
}}
Here $U$ is a unitary matrix in color space which is defined on the
links of lattice, and can be thought of as the path-ordered exponential
of the gluon field,
$$
U\sim P \exp i g \int A_\mu d x^\mu.
$$
$W$ satisfies the discretized version of the Green's function equation
$$
\left(v\cdot D\right) W(\bfu x,t; \bfu y, s) = {1+\gamma^0\over2}
\ \delta(\bfu x - \bfu y) \delta(t-s),
$$
\eqn\defeqn{
{1\over a}\left[W(\bfu x,t; \bfu y, s)-
U(\bfu x, t; \bfu x, t-a) W(\bfu x,t-a; \bfu y, s)\right]
= a^{-4}\ {1+\gamma^0\over2}\ \delta_{\bfu x \bfu y}\, \delta_{t s},
}
where $\delta_{ab}$ is a Kronecker delta. To avoid complicating the
notation in the correlation functions to be computed below, $(\bfu x,t)$
will be denoted by $x$.

Consider the meson correlation function (at $\bfu y = \bfu x$, $y^0> x^0$)
\eqn\hmeson{
\vev {\bar q\, \Gamma^\dagger Q_v\left(y\right)\
\bar Q_v \Gamma\, q\left(x\right) }= (-1) \int d\mu
\Tr W(y,x)\, \Gamma\, S(x,y;m_q)\, \Gamma^\dagger,
}
where the trace is over spinor and color indices, the minus sign is due
to Fermi statistics, and the fermion propagator for the quark $q$ is
\eqn\lprop{
S(x,y;m_q) = \bra{x}\left(\dsl+m_q\right)^{-1}\ket{y}.
}
The Euclidean measure $d\mu$ in the functional integral is defined using
the QCD action of the light quarks and gluons,  and angle brackets are
used to denote the functional integral over the light quark and gluon fields,
$$
\vev{f} \equiv \int d \mu\  f = {1\over Z}\int \CD A
\ e^{-S_{\rm eff}(A)}\ f,
$$
where
$$
S_{\rm eff} = {1\over 2}\Tr G^2 - \sum_i\ln \det \left(\dsl +
m_i\right),
$$
and the sum is over the different light quark flavors. The QCD Euclidean
path integral measure is real and positive for $\theta_{\rm QCD}=0$,
since QCD is a vector-like gauge theory, so one can use the
Cauchy-Schwarz inequality on eq.~\hmeson,
\eqn\mineq{\eqalign{
\abs{\vev {\bar q\, \Gamma^\dagger\, Q_v\left(y\right)\
\bar Q_v\, \Gamma\, q\left(x\right) }}^2 &\le
\int d\mu \Tr\, \Gamma\, \Gamma^\dagger\ \Tr\, W(y,x)\, W(y,x)^\dagger \cr
&\qquad\times\int d\mu\Tr S(x,y;m_q)\, S^\dagger(x,y;m_q).
}}
The term on the r.h.s.~proportional to the light quark propagator can be
reexpressed as a meson correlation function (neglecting annihilation
graphs), since
\eqn\picorr{
\vev{\bar q\, i\gamma_5\, q\left(y\right)\  \bar q\, i\gamma_5\, q
\left(x\right)} = \int d\mu \Tr S(x,y;m_q)\, S^\dagger(x,y;m_q),
}
using
$$
\gamma_5\ S(y,x;m_q)\ \gamma_5 = S(x,y;m_q)^\dagger.
$$
Combining eqs.~\mineq\ and \picorr, and using ($N_c$ is the number of colors)
\eqn\wineq{
\int d\mu \Tr\, \Gamma\, \Gamma^\dagger\ \Tr\, W(x,y)\ W(x,y)^\dagger
\le 2 N_c \Tr\, \Gamma\, \Gamma^\dagger\ a^{-6} \int d \mu
}
which follows from eq.~\eprop\ and the unitarity of $U$ gives
\eqn\mesonineq{
\abs{\vev {\bar q\, \Gamma^\dagger\, Q_v\left(y\right)\
\bar Q_v\, \Gamma\, q\left(x\right) }}^2\le C\
\vev{\bar q\, i\gamma_5\, q\left(x\right)\
\bar q\, i\gamma_5\, q\left(y\right)}
}
where $C$ is a (cutoff dependent) constant which is independent of $x$
and $y$. At large distances, the Euclidean correlation function is
dominated by the lightest particle in a given channel. For the heavy
quark effective theory, this is true for the correlation function in
eq.~\mineq\ provided $\bfu x = \bfu y$ (since we are working in the
frame where $\bfu v = 0$), otherwise the correlation function vanishes.
The long distance behavior of the correlation function of any operator is
$$
\vev{ O\left(t,\bfu x\right) O^\dagger\left(s,\bfu x\right)}
\sim A\ e^{-M\left( t - s \right)},
$$
where $M$ is the mass of the lightest state that can be created from the
vacuum by $O$, and $t>s$. The lightest state that can couple to the
l.h.s.~of eq.~\mesonineq\ is a heavy quark meson with the spin quantum
numbers of $\Gamma$, \eg\ for $\Gamma=i\gamma_5$, it is the lightest
pseudoscalar, \etc\ On the r.h.s., the state is the lightest
pseudoscalar $q\bar q$ state (neglecting annihilation diagrams), so we
get the inequality
\eqn\Mresult{
m\left(\bar Q\,\Gamma\, q\right) - m_Q\ \ge
\ \half\ m'\left(\,\bar q\, i\gamma_5\, q\,\right),
}
where we have replaced the mass of the heavy meson in the effective
theory by the difference of the mass of the heavy meson and heavy quark.
The prime on $m$ is a reminder that annihilation graphs are neglected.

The inequality eq.~\Mresult\ is stronger than the inequality \witten\nussinov
\eqn\weakresult{
m\left(\bar Q\,\Gamma\, q\right) \ \ge
\ \half\  m'\left(\bar Q\,i\gamma_5\, Q\right)+
\half\ m'\left(\,\bar q\, i\gamma_5\, q\,\right),
}
which can be derived in QCD without assuming that $Q$ is infinitely
heavy.  The inequality eq.~\Mresult\ is an inequality between two
quantities which are finite in the infinite mass limit. Eq.~\weakresult\
can be rewritten in the form
$$
m\left(\bar Q\,\Gamma\, q\right) -m_Q\ \ge
\ \half\  m'\left(\bar Q\,i\gamma_5\, Q\right) -m_Q+
\half\ m'\left(\,\bar q\, i\gamma_5\, q\,\right).
$$
In the limit $m_Q\rightarrow\infty$, the mass of the lightest $\bar
Q\,i\gamma_5\, Q$ bound  state can be computed using QCD perturbation
theory to be $2m_Q-E_b$, where the binding energy $E_b$ is $E_b= \half
\left[\frac43 \alpha_s \left( m_Q\right) \right]^2 m_Q$ to lowest order
in perturbation theory. $E_b\rightarrow\infty$ in the heavy quark limit,
so eq.~\weakresult\ becomes the trivial inequality ${\rm finite} >
-\infty$ in the heavy quark limit.

The binding energy is missing in the inequality eq.~\Mresult, which
makes it a stronger inequality than eq.~\weakresult\ in the heavy quark
limit. The reason the term is missing in eq.~\Mresult\ is that the heavy
quark propagator in the effective theory is a straight line in
space-time with v=(1,0,0,0), and no quantum fluctuations. The r.h.s.~of
the Cauchy-Schwarz inequality in eq.~\mineq\ contains the heavy quark
propagator and its hermitian conjugate, which is the propagator for an
anti-quark. The two propagators represent the interaction of a
quark-antiquark pair at the same point in space, since the heavy quark
does not move. This static $Q\bar Q$ pair is color neutral, and does not
interact with the gauge field (equivalently $WW^\dagger=1$ is
independent of the gluon vector potential). On the other hand, quantum
fluctuations are included if one works with finite quark masses in the
full QCD theory, and the propagator $S$ is obtained by summing over all
paths that start at the initial point $x$, and end at the final point
$y$. Thus $\abs{S}^2$ is obtained by summing over all paths for a
quark-antiquark pair to start at $x$ and end at $y$, with {\sl
independent} paths for the quark and antiquark. There are terms in the
sum where the quark and antiquark paths are not identical. These paths
can exchange a gluon between them,  and this gives the binding energy
$E_b$. The divergent binding energy $E_b$ is the reason the heavy quark
effective theory breaks down for hadrons containing more than one heavy
quark. The QCD inequalities follow from studying the square of the
original meson correlation function, and thus have the particle content
of a meson-antimeson pair. In deriving eq.~\Mresult, the correlation
function was computed in the effective theory {\sl before} squaring. In
eq.~\weakresult, the square of the correlation function was computed in
the full theory before taking the $m_Q\rightarrow\infty$ limit. This
requires the study of the correlation function for $Q\bar Q q\bar q$
which contains two heavy quarks, so the heavy quark effective theory
breaks down. The order of limits does not commute, which is why we
obtain a stronger inequality by computing correlation functions directly
in the effective theory, and then using the Cauchy-Schwarz inequality.

The l.h.s.~of eq.~\Mresult\ (in the pseudoscalar channel) is the meson
$\bar \Lambda$ parameter of the heavy quark effective theory, which is a
physical parameter that occurs in the corrections to meson form factors
\ref\luke{M.E. Luke, \pl{B252}{1990}{447}}\ref\ggw{H. Georgi, B.
Grinstein, and M.B. Wise, \pl{B252}{1990}{456}}\fln,
so the inequality can be rewritten as
\eqn\mresult{
2 \,\bar\Lambda\left(\bar Q\,\Gamma\, q\right) \ \ge
\ m'(\,\bar q\, i\gamma_5\, q\,).
}
The inequality determines the sign of $\bar\Lambda$ to be positive. The
r.h.s.~of the inequality can also be determined, since the $\bar q\,
i\gamma_5\, q$ mass neglecting annihilation graphs is the same as the
$\bar q_1\, i\gamma_5\, q_2$ mass in the limit that $q_1$ and $q_2$ have
equal masses. The masses of these states can be derived using chiral
perturbation theory, and the known values of the pion and kaon masses,
\eqn\mprimes{\eqalign{
m'(\,\bar u\, i\gamma_5\, u\,) &= \sqrt{2 m_u\over m_u+m_d}\ m_{\pi^0} =
114\ \MeV,\cr
\noalign{\medskip}
m'(\,\bar d\, i\gamma_5\, d\,) &= \sqrt{2 m_d\over m_u+m_d}\ m_{\pi^0}=
153\ \MeV,\cr
\noalign{\medskip}
m'(\,\bar s\, i\gamma_5\, s\,) &= \sqrt{2 m_s\over m_u+m_d}\ m_{\pi^0}=
685 \ \MeV,\cr
}}
using the lowest order values for the light quark mass ratios,
$m_u/m_d=0.56$, and $m_s/m_d=20.1$ \ref\weinberg{S. Weinberg, Trans.
N.Y. Acad. Sci. 38 (1977) 185}. One can improve the values of the meson
masses by using the second order formul\ae\ for the pseudoscalar masses
\ref\second{J. Gasser and H. Leutwyler, Ann.~Phys.~ 158 (1984) 142, \np
{250} {1985} {465}\semi D.B. Kaplan and A.V. Manohar, \prl{56} {1986} {2004}}.
This will be discussed elsewhere. Eqs.~\Mresult\ and \mprimes\ gives the
inequalities for the meson $\bar\Lambda$ parameters,
\eqn\Qineq{\eqalign{
\bar\Lambda\left(Q\bar u\right) &\ge 57\ \MeV,\cr
\bar\Lambda\left(Q\bar d\right) &\ge 76\ \MeV,\cr
\bar\Lambda\left(Q\bar s\right) &\ge 343\ \MeV.\cr
}}

Mass inequalities for baryon containing one heavy quark are obtained by
studying the $Qq_1q_2$ correlation function. There are no new
subtleties which occur in this case, so we will not present the details
here. The QCD inequality is
\eqn\baryonineq{\eqalign{
&\abs{\vev{Qq_1q_2\left(y\right) \bar Q\bar q_1\bar q_2\left(x\right)}}\cr
&\le \vev{ \left[ \Tr\,S(x,y,m_1)\,S^\dagger(x,y,m_1)\
 \Tr\,S(x,y,m_2)\,S^\dagger(x,y,m_2)\right]^{1/2}}\cr
&\le \vev{  \Tr\,S(x,y,m_1)\,S^\dagger(x,y,m_1)}^{1/2}
\vev{ \Tr\,S(x,y,m_2)\,S^\dagger(x,y,m_2)}^{1/2},
}}
where the $Qq_1q_2$ baryon on the l.h.s.~can be in any spin channel, and
$m_1$ and $m_2$ are the masses of $q_1$ and $q_2$ respectively.
Comparing the exponential fall-off of both sides gives the inequality for the
$\bar\Lambda$ parameter for baryons containing a heavy quark,
\eqn\bresult{
\bar\Lambda\left(Qq_1q_2\right) = m\left(Qq_1q_2\right)-m_Q
\ge \half\, m'\left(\bar q_1 \,i\gamma_5\,q_1\right)+
\half\, m'\left(\bar q_2\,i\gamma_5\, q_2\right),
}
where the prime signifies that annihilation graphs are neglected for the
mesons on the r.h.s.~of the inequality. This gives the inequalities
\eqn\Bineq{\eqalign{
&\bar\Lambda\left(Qss\right) \ge 685\ \MeV,\cr
\noalign{\smallskip}
&\bar\Lambda\left(Qus\right) \ge 400\ \MeV,\cr
\noalign{\smallskip}
&\bar\Lambda\left(Qds\right) \ge 419\ \MeV,\cr
}\qquad\eqalign{
&\bar\Lambda\left(Quu\right) \ge 114\ \MeV,\cr
\noalign{\smallskip}
&\bar\Lambda\left(Qud\right) \ge 133\ \MeV,\cr
\noalign{\smallskip}
&\bar\Lambda\left(Qdd\right) \ge 153\ \MeV,\cr
}}
using eq.~\mprimes.

The l.h.s.~of the inequalities eqs.~\mresult\ and \bresult\ are finite
in the limit that the light quark mass $m_q\rightarrow0$. $SU(3)$
symmetry breaking corrections to the $\bar\Lambda$ parameter are of
order $\Delta m_q/ \Lambda_\chi$, where $\Delta m_q$ is the
difference of two light quark masses, and $\Lambda_\chi$ is the chiral
symmetry breaking scale which is of order 1~GeV. The r.h.s.~of
eqs.~\mresult\ and \bresult\ vanish as $m_q^{1/2}$ as $m_q\rightarrow0$.
Thus the strongest inequality for hadrons containing $u$ and $d$ quarks
is obtained by using $SU(3)$ symmetry and the inequality for hadrons
containing $s$ quarks. This procedure has the disadvantage that the
resulting inequalities are only approximate, since $SU(3)$ breaking
corrections can be either positive or negative. A better method is to
use the inequalities for hadrons containing a $s$ quark and the
experimental values for the hadron mass differences to obtain
inequalities for hadrons containing $u$ and $d$ quarks. The inequalities
on $\bar\Lambda$ obtained by this method are stronger than the
inequalities in eq.~\Qineq\ because we have included some extra
information from experiment. For the $c$ quark system, one has the inequalities
\eqn\cineq{
m_c = m(D_s)-\bar\Lambda(D_s) \le m(D_s)-343\ \MeV \le 1627 \ \MeV,
}
using $\bar\Lambda(D_s)\ge343$~MeV from eq.~\Qineq, and
\eqn\cresults{\eqalign{
\bar\Lambda(D^0)&= \bar\Lambda(D_s) + m(D^0)-m(D_s) \ge 237\ \MeV,\cr
\bar\Lambda(D^+)&= \bar\Lambda(D_s) + m(D^+)-m(D_s) \ge 243\ \MeV,\cr
\bar\Lambda(\Lambda_c^+)&=\bar\Lambda(D_s)+m(\Lambda_c^+)-m(D_s) \ge 657\ \MeV,
}}
and similarly for the other hadrons containing a $c$ quark. The
experimental numbers have been taken from the 1992 Particle Data Book
\ref\pdg{Review of Particle Properties, K. Hikasa, \etal, \physrev {D45}
{1992} {1}}, and we have used $1\sigma$ experimental errors.
A similar analysis for hadrons containing a $b$ quark gives
\eqn\bineq{
m_b \le m(B_s)-343\ \MeV \le 5068 \ \MeV,
}
using the experimental value $80\le m(B_s)-m(B)\le 130$~MeV from \pdg, and
$\bar\Lambda(B_s)\ge343$~MeV from eq.~\Qineq. There are also inequalities
for the $\bar\Lambda$ parameters,
\eqn\bresults{\eqalign{
\bar\Lambda(B^{0,+}) &= \bar\Lambda(B_s) + m(B^{0,+})-m(B_s) \ge 213\ \MeV,\cr
\bar\Lambda(\Lambda_b^0)&=\bar\Lambda(B_s)+m(\Lambda_b^0)-m(B_s) \ge 523\ \MeV.
}}
The inequalities for the $\bar\Lambda$ parameters for the $b$ system are
weaker than those for the $c$ system because of the larger experimental
uncertainties in the masses of the $B_s$ and $\Lambda_b^0$ hadrons. In the
heavy quark limit, the $\bar\Lambda$ parameters for the $c$ and $b$
hadrons are the same, and one can use the inequalities eq.~\cresults\
for the $b$ system. Finally, the inequalities eqs.~\cineq--\bresults\
imply that $\bar\Lambda/m_Q$ which characterizes the size of $1/m_Q$
corrections is greater than 0.15 for $D$ mesons, 0.04 for $B$ mesons,
0.4 for the $\Lambda_c^+$, and 0.1 for the $\Lambda_b^0$.

We would like to thank U. Wolff for helpful discussions.
This work was supported  in part by the U.S.~Department of Energy
under Grant No.~DOE-FG03-90ER40546 and by a
NSF Presidential Young Investigator Award PHY-8958081.

\listrefs
\bye